\documentclass[preprint,12pt, times]{elsarticle}

\usepackage{amssymb}

\usepackage{graphicx}
\usepackage{subcaption}
\usepackage{amsmath}
\usepackage{booktabs}
\usepackage{multirow}
\usepackage{verbatim}

\providecommand{\abs}[1]{\lvert#1\rvert}
\DeclareMathOperator{\sgn}{sgn}
\begin{document}

\begin{frontmatter}



\title{Inverse NN Modelling of a Piezoelectric Stage with Dominant Variable}


\author[CDU]{Gangfeng YAN}
\corref{YGF}
\ead{yangangfeng@cdu.edu.cn}
\address[CDU]{College of Information Science and Engineering, Chengdu University, Chengdu, 610106, Sichuan, China}

\cortext[YGF]{Corresponding author}

\author[NUS]{Hang Jian SOO}
\ead{e0153346@u.nus.edu}

\author[NUIS]{Khalid ABIDI}
\ead{Khalid.Abidi@newcastle.ac.uk}
\address[NUIS]{Newcastle University Singapore, SIT Building $@$ Nanyang Polytechnic
172A Ang~Mo~Kio Avenue 8 \#05-01
Singapore 567739}

\author[NUS]{Jian-Xin XU}
\ead{elexujx@nus.edu.sg}
\address[NUS]{Department of Electrical and Computer Engineering, National University of Singapore, Singapore 117576 }

\begin{abstract}
This paper presents an approach for developing a neural network inverse model of a piezoelectric positioning stage, which exhibits rate-dependent, asymmetric hysteresis. It is shown that using both the velocity and the acceleration as inputs results in overfitting. To overcome this, a rough analytical model of the actuator is derived and by measuring its response to excitation, the velocity signal is identified as the dominant variable. By setting the input space of the neural network to only the dominant variable, an inverse model with good predictive ability is obtained. Training of the network is accomplished using the Levenberg-Marquardt algorithm. Finally, the effectiveness of the proposed approach is experimentally demonstrated.
\end{abstract}

\begin{keyword}
Neural network modelling
\sep Dominant variable
\sep Complex nonlinearity
\sep Piezoelectric positioning stage
\sep Model-based control



\end{keyword}

\end{frontmatter}



\pagestyle{myheadings}
\markright{Piezoelectric Stage Inverse Modelling}

\renewcommand{\vec}[1]{\mathbf{\MakeLowercase{#1}}}
\newcommand{\mat}[1]{\mathbf{\MakeUppercase{#1}}}

\section{Introduction}
Piezoelectric actuators are an attractive choice for high precision micro-positioning applications due to their quick response time, large force output, and high bandwidth. Example application areas include atomic force microscopes \cite{chang1999,croft2000}, adaptive optics \cite{song2009}, and computer components \cite{yang2010}.

The inherently nonlinear nature of piezoelectric actuators means that their full capabilities can only be realised with careful control, but this is challenging due to the complexity of the nonlinearity, which is a combination of hysteresis and creep phenomena. Nevertheless, many advanced control methods have been successfully applied to piezoelectric actuators (see \cite{xu2016}). The choice of method certainly depends on the application. For example, Iterative Learning Control is suitable for repetitive motions \cite{abidi2011}, while the robustness of Sliding Mode Control is effective in counteracting exogeneous disturbances \cite{abidi2009}.

On the other hand, the standard control strategy of combining feedforward and feedback control is simple and also feasible, but to attain high precision an accurate actuator model (or its inverse) is needed for feedforward compensation. As hysteresis is the dominant source of error in piezoelectric actuators (the creep effect occurs over a longer timescale), much work has focused on modelling it accurately. 

The key issue in modelling hysteresis is how to adequately describe its dynamical nature. In other words, knowledge of the current input is insufficient; some idea of its past is necessary as well. In a sense, this is a problem of transforming a multi-valued mapping into a one-to-one mapping. Matters are complicated by the asymmetry and rate-dependence of the hysteresis.

Since physics-based models are hard to derive and only narrowly applicable, most available models are phenomenological. The applicability of phenomenological models across different physical systems, such as in the magnetisation of materials, and the displacements generated by shape memory alloys, make it possible to draw from hysteresis modelling work in those areas. However, parameter estimation for phenomenological models may be less straightforward compared to physical models, because parameters do not directly correspond to physically measurable quantities. 

Hassani et al give a recent review \cite{hassani2014} of many analytical models of hysteresis, which they broadly categorise as involving either the use of hysteretic operators, or differential equations. The classic Preisach model, and the Krasnoselskii-Pokrovskii model, are representative of the operator approach, where complicated hysteresis models are constructed by the superposition of elementary hysteresis operators. Using more of these building blocks increases the precision of the model, but also the number of parameters that need to be estimated. 

For the purpose of feedforward compensation an inverse hysteresis model is typically required. The Preisach and Krasnoselskii-Pokrovskii models are not amenable to analytic inversion, although numeric inversion is possible \cite{song2017}. The Prandtl-Ishlinskii (P-I) model overcomes this difficulty, and its inverse has been successfully employed for feedforward compensation in a piezoelectric actuator \cite{aljanaideh2011}. 

Since hysteretic behaviour depends on the past state of the system, it is natural to consider the use of differential equations for modelling hysteresis. The Buoc-Wen model is one such model which has been applied to piezoelectric actuators \cite{xiao2014}. It is readily incorporated into the state space model of an actuator, and due to its mathematical form can be directly used for compensation without inversion \cite{rakotondrabe2011}.

Classic implementations of the abovementioned hysteresis models do not account for asymmetry and rate-dependence. Many variants of these models (especially of the P-I variety) have thus been developed to address these shortcomings, for example in \cite{xiao2014, ang2007, gu2014, li2014}. The necessity of such modifications suggests an important limitation of analytical phenomenological models: the form of an analytical expression often contains implicit assumptions about the characteristics of the hysteresis which may not reflect reality. Moreover, numerical issues can arise during computation of these analytical models (e.g. Tan et al \cite{tan2009} show that the inverse of the P-I model can become numerically ill-conditioned.) To address these limitations, a black-box modelling method is desirable to avoid making assumptions about the hysteresis or any other physical effects that may be present. Neural networks are a popular example, and have been successfully applied to many nonlinear system identification problems.

Much work has been reported on the use of neural networks to build hysteresis models. Most approaches employ a static, feedforward neural network with a variety of activation functions and learning algorithms; however, such neural networks are alone incapable of describing dynamical behaviour like hysteresis. Therefore in these approaches another analytic mapping is used to extract information about the state of the hysteresis from its input/output. The output(s) from this mapping (also called a `hysteretic operator' by various authors), along with the input to the hysteresis, are fed into a neural network which maps to the output of the hysteresis model. Thus, the neural network is said to map from an `expanded input space'. Examples of this approach can be seen in the work of Zhao and Tan \cite{zhao2006}, Li and Tan \cite{li2004}, and Ma et al, who have proposed a variety of hysteretic operators \cite{ma2008, ma2011, ma2014, ma2016}. 

More elaborate variations of this approach have also been developed to improve accuracy, especially in accounting for rate-dependence. The model of Zhang and Tan combines the aforementioned approach with a superposition of first-order differential operators \cite{zhang2010}. In the work of Dang and Tan \cite{dang2005}, the mapping computes an inner product using the input and delayed output of the hysteresis; furthermore, the neural network's expanded input space includes as a third input, the delayed output of the hysteresis. The same authors also consider a similar arrangement with a simpler mapping in \cite{dang2005NARX}, which is inspired by conceiving the problem as finding a nonlinear auto-regressive model with exogeneous input (NARX). In the work of Dong et al \cite{dong2008}, the mapping computes a `generalised gradient' of the hysteresis; the expanded input space is similar to Dang and Tan's work but with the further addition of a fourth input, \emph{viz.} the time derivative of the hysteresis input.

The above neural network-based models are not truly `black-box' based because the formulation of the mappings or hysteretic operators involves making some assumptions about the shape of the hysteresis. It would be preferable to do away with these complicated mappings and completely rely on an appropriate neural network architecture to directly model the dynamics of the hysteresis. In the model of Li and Chen \cite{li2013}, which was conceived as a nonlinear auto-regressive moving average model with exogeneous input (NARMAX), a history of past input and output values to the hysteresis is stored and fed to the input of a static feedforward neural network. Mai et al \cite{mai2016} propose the use of a time-delay dynamic neural network, which processes inputs subjected to some time-delay, to model hysteresis in shape memory alloys. Experiments showed that the time-delay could be optimised to minimise the modelling error (interestingly, \cite{ai2016} also shows another example where introduction of a time delay can be beneficial). Liu and Xiu \cite{liu2007} use neurons with a hysteretic activation function and backpropagation learning. Deng et al \cite{deng2016} use a modified Elman network with the Levenberg-Marquardt algorithm. Laudani et al \cite{laudani2016} use a Fully Connected Cascade network architecture with neuron-by-neuron learning to model magnetisation of materials.

Other less common approaches have also been investigated. Examples are the work by Deng and Tan \cite{deng2009}, who use a hysteretic operator in a NARMAX model and a recursive least squares method to determine the model parameters, and the work by Xu \cite{xu2013} which employs a Least Squares Support Vector Machine to build the model.

In this work, a method for inverse modelling of a piezoelectric positioning stage is described. The stage is actuated by a piezoelectric motor which enables extremely precise positioning, but unlike standard stack or bender piezoelectric actuators, the range of travel is very much greater, on the order of centimeters. This also means that frictional effects could be very significant. Unlike most neural network hysteresis models described above, a simple static feedforward neural network is used here, with no need for complicated hysteresis operators. Yet it is capable of modelling the second-order dynamics of the actuator, by taking as inputs the first and second time derivatives of the position i.e. velocity and acceleration. 

However, through experiment it was found that if both these variables are input to the neural network (as would typically be the case), then overfitting would occur and the resulting model would be inadequate for the purpose of prediction. Instead, by restricting the neural network's input to only the dominant variable of the piezoelectric positioning stage (\emph{viz.} the velocity), this yields an inverse model with good prediction performance.

This paper is organised as follows. A brief description of the piezoelectric positioning stage and experimental setup is presented in Section~\ref{sect:exptSetup}, and its nonlinear characteristics are investigated in Section~\ref{sect:nonlinearCharacteristics}. Construction and performance evaluation of a backpropagation neural network inverse model using the typical approach is described in Section~\ref{sect:NNtypical}. In Section~\ref{sect:NNDominantVariable} the neural network is designed using the dominant variable approach instead. Results demonstrating the effectiveness of this approach are presented in Section~\ref{sect:predictionPerformance}.

\section{Experimental Setup}
\label{sect:exptSetup}
The piezoelectric positioning stage investigated in this work is a PLS8 manufactured by PBA Systems. The maximum range of travel of the stage is 115 mm, and the maximum velocity is 230 mm/s. 

The stage is actuated by a Nanomotion HR-8 piezoelectric motor. Its working principle is briefly described (c.f. Fig. \ref{fig:workingPrinciple}):
The actuating elements are a set of piezoelectric ceramic fingers. The fingertips, protruding from one end of the motor, are mounted in compression against the drive belt of the work platform. When driven by electrical signals from the motor driver, ultrasonic standing waves are produced and the high frequency longitudinal extension and lateral bending of the finger generates an elliptical motion at the fingertips. The force exerted on the drive belt by the fingertips moving in such a manner produces linear motion along the direction as shown in Fig. \ref{fig:workingPrinciple}. The control voltage applied to the motor driver determines the velocity of motion. In the absence of drive voltage input, the pressure of the ceramic fingertips on the drive belt maintains a holding torque on the work platform.

\begin{figure}
\centering
\includegraphics[width=0.7\textwidth]{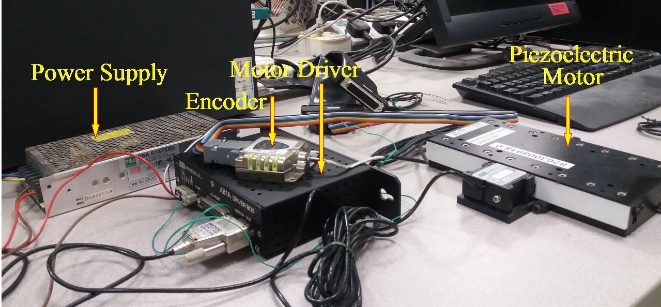}
\caption{The piezoelectric positioning stage system.}
\label{fig:piezoStagePhoto}
\end{figure}

\begin{figure}
\centering
\includegraphics[width=0.7\textwidth]{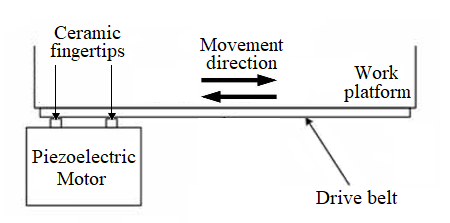}
\caption{Schematic illustrating the working principle of the piezoelectric positioning stage.}
\label{fig:workingPrinciple}
\end{figure}

Position is measured by a Mercury 3000 encoder made by Celera Motion while the velocity and acceleration signals are obtained by numerical differentiation of the position.
All control and measurement algorithms are implemented with MATLAB/SIMULINK on a host computer, and executed by a dSPACE DS1104 card installed inside. Signal acquisition and generation are respectively via the DS1104’s 12-bit Analog-to-Digital Converter (ADC) channels (800 ns conversion time) and 16-bit Digital-to-Analog converter (DAC) channels (10 µs settling time), both having ±10 V dynamic range. These channels interface with the piezoelectric motor driver and the encoder. Through a user interface on the dSPACE ControlDesk software, experiments are performed with parameter adjustments and measurements made in real time. The entire piezoelectric positioning stage system is shown in Fig.~\ref{fig:piezoStagePhoto}.


\section{Nonlinear Characteristics of the Piezoelectric Positioning Stage}
\label{sect:nonlinearCharacteristics}

The complexity of the nonlinear relationship between applied voltage and the resulting motion of the piezoelectric positioning stage is investigated by conducting a number of experiments. 

\begin{figure}
\centering
\includegraphics[width=0.7\textwidth]{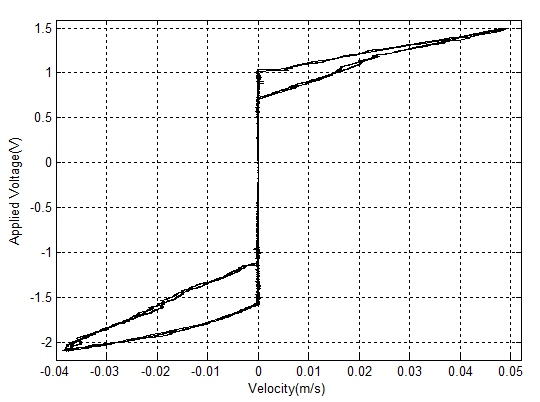}
\caption{Relationship between applied voltage/frictional force and velocity of the positioning stage.}
\label{fig:forceVsVelocityHysteresis}
\end{figure}

Essential features of the voltage-velocity relationship are observed by applying a triangle waveform to the input (Fig.~\ref{fig:forceVsVelocityHysteresis}). In addition, by using a low slew rate to move the stage at low speed with minimal acceleration, it becomes reasonable to assume that the actuation force (as reflected by the applied voltage) is applied solely to overcome the frictional forces. Therefore, the relationship between frictional forces and velocity can also be inferred from Fig.~\ref{fig:forceVsVelocityHysteresis}. Apart from the presence of hysteresis, this result also suggests that Coulomb friction and viscous friction forces are significant and need to be accounted for in any model of the system. Moreover, frictional characteristics are dependent on the direction of motion.

\begin{figure}
\centering
\begin{subfigure}{0.5\textwidth}
\includegraphics[width=\textwidth]{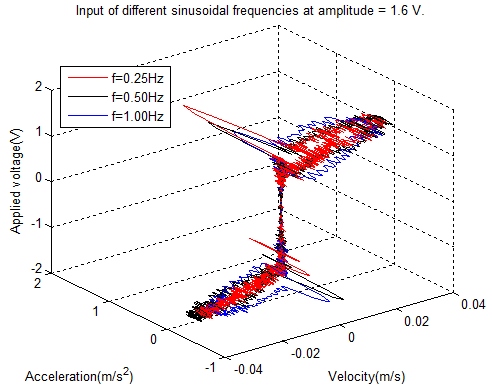}
\end{subfigure}
\begin{subfigure}{0.48\textwidth}
\includegraphics[width=\textwidth]{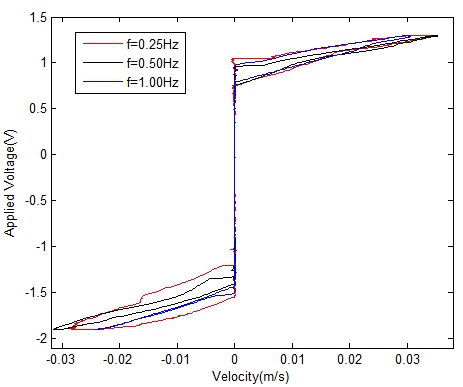}
\end{subfigure}
\caption{Velocity and acceleration in response to sinusoidal input with amplitude 1.6V, at varying frequencies.}
\label{fig:nonlinearitiesVsFreq}
\end{figure}

\begin{figure}
\centering
\begin{subfigure}{0.5\textwidth}
\includegraphics[width=\textwidth]{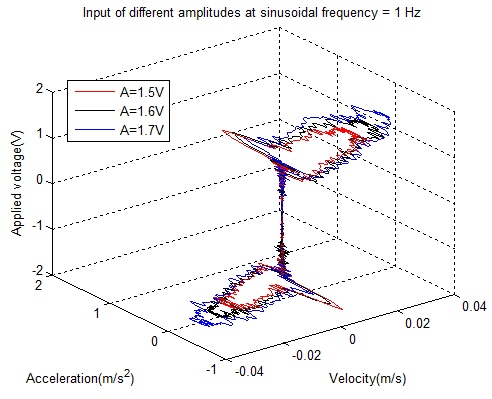}
\end{subfigure}
\begin{subfigure}{0.48\textwidth}
\includegraphics[width=\textwidth]{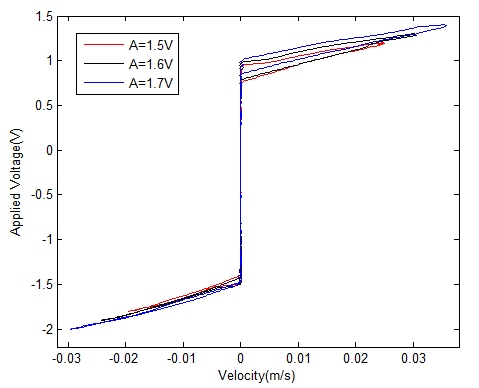}
\end{subfigure}
\caption{Velocity and acceleration in response to 1Hz sinusoidal input at varying amplitudes.}
\label{fig:nonlinearitiesVsAmpl}
\end{figure}

Further complexity is revealed by applying sinusoidal signals described by the function $v = A \sin( 2\pi ft - \pi/2)- 0.3$, for various values of amplitude $A$ and frequency $f$. The response of the system (velocity and acceleration) are shown in Fig.~\ref{fig:nonlinearitiesVsFreq} and Fig.~\ref{fig:nonlinearitiesVsAmpl}. The results show the effects of varying frequency at a constant amplitude, and the effects of varying amplitude at a constant frequency.

From the figures it is evident that the asymmetrical hysteresis shown in Fig.~\ref{fig:forceVsVelocityHysteresis} is a basic qualitative feature of the voltage-velocity/acceleration relationships. However, the specific shapes of the curves depend on the temporal variation of the applied voltage, in a complicated and rather irregular manner. These results suggest that the nonlinear behaviour of the piezoelectric stage cannot be easily characterised by analytical modelling.

\section{Neural Network Modelling -- Typical Approach}
\label{sect:NNtypical}
Amongst the variety of neural networks applied in various fields of engineering, perhaps the most well-known is the feedforward neural network using the Back-Propagation (BP) learning algorithm (i.e. the BP neural network), which is capable of approximating any continuous nonlinear function, and is a suitable candidate for constructing an inverse model of the piezoelectric positioning stage. 

\subsection{Backpropagation Neural Networks}
A BP neural network is a network of interconnected processing elements (i.e. neurons). Each neuron computes a weighted sum of its inputs $u$, which is then mapped through an activation function $f$ to generate an output $O$. In this work, the hyperbolic tangent function was chosen as the activation function. Neurons are arranged in layers which are connected in cascade, such that the output of every neuron in a given layer is connected to the input of every neuron in the next (i.e. a feedforward network). The entire network comprises an input layer, an output layer, and one or more hidden layers in between. More precisely, a neuron $j$ in the layer $k$ computes the function given as
\begin{align}
	u^k_j &= \sum_{i} w_{ji} O_{i}^{k-1}	\\
	O_j^k &= f(u^k_j) = \frac{e^{u^k_j} - e^{-u^k_j}}{e^{u^k_j} + e^{-u^k_j}}
\end{align}
where $w_{ji}$ is the weight applied to $O_i^{k-1}$, the output of the $i$-th neuron in the preceding layer $(k-1)$. The outputs of the input layer $O_i^0$ are simply the input values presented to the network.
Thus a function is computed by the network as input signals propagate through and are transformed by successive layers from input to output. At the output layer, the output error of each neuron $j$, in response to a given input pattern, is the difference between the desired output $D_j$ and its actual output $O_j$. 
During the training process, these errors are propagated back through the network in the opposite direction, and weights are adjusted accordingly in order to minimise the \emph{overall error} of the network. A suitable measure of this is given by the instantaneous error energy $e$, which is a sum of squared errors of all output neurons for some input pattern given as
\begin{equation}
	e = \frac{1}{2} \sum_j(D_j - O_j)^2
\end{equation}
In this work, the Levenberg-Marquardt training algorithm (a variant of classic backpropagation) is employed, with a momentum term to speed up convergence. Weights are updated at the end of every training epoch, according to the rule given by
\begin{equation}
	\vec{w}_{n+1} = \vec{w}_n - (\mat{J}_n^T \mat{J}_n + \mu \mat{I})^{-1} \mat{J}_n^T \vec{e}_n + \alpha \Delta \vec{w}_n
\end{equation}

where $n$ refers to the training epoch/iteration, $\vec{w}$ is a vector of all the weights in the network, $\vec{e}$ is a vector of output errors, $\mat{J}$ is the Jacobian matrix of the errors with respect to the weights in the network, $\mat{I}$ is the identity matrix, $\mu$ is the combination coefficient and $\alpha$ is the momentum coefficient.

The operation of this algorithm is understood as an adaptive blending of the steepest gradient descent method and the Gauss-Newton method. This is controlled by the value of the combination coefficient $\mu$ which is adjusted every training iteration in order to achieve fast and stable convergence to the error minimum. Initially $\mu$ is large and the steepest descent method dominates which gives the advantage of speed; as the error minimum is approached, $\mu$ is progressively reduced and the Gauss-Newton algorithm dominates. 

Using the typical approach, a BP neural network was constructed with velocity and acceleration as the input, and applied voltage as the output. It was decided to use a single hidden layer; this is premised on the well-known result that one hidden layer is sufficient to approximate any continuous function to any desired level of accuracy by selecting a suitable number of neurons. From experimental observation it was found that a hidden layer with 120 neurons was adequate.

\subsection{Training and Performance}
\label{sect:trngPerformance}
The datasets used to train and test the network are sequences of velocity and acceleration values (the inputs), paired with the corresponding control voltage values (the output). 

In order to collect data, the piezoelectric positioning stage is made to roughly track a sinusoidal position reference signal, using a proportional feedback position controller. At each sampling instant, the control voltage applied by the controller and the position of the stage measured by the encoder are recorded. Numerical differentiation of the position gives the corresponding velocity and acceleration values. 

This process is repeated with the position reference signal set to various amplitude and frequency values. These parameter combinations are selected to span the region of operation of the piezoelectric stage, which is bounded by manufacturer-specified limits on the range of motion and operating speed. 

\begin{figure}
\centering
\includegraphics[width=0.7\textwidth]{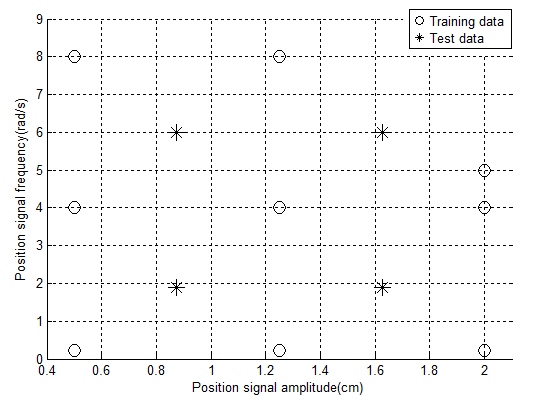}
\caption{Parameter combinations used to construct the training and test data sets.}
\label{fig:NNdataSets}
\end{figure}

Data from two different sets of amplitude-frequency parameters is used to generate the training and test datasets, as indicated in Fig.~\ref{fig:NNdataSets}. A dataset is generated by concatenating one segment of data from every parameter combination used, in random order. Each segment corresponds to one cycle of the sinusoidal position reference signal, and is chosen to avoid introducing jump discontinuities when concatenated with adjacent segments.

\begin{figure}
\centering
\begin{subfigure}{0.45\textwidth}
\includegraphics[width=\textwidth]{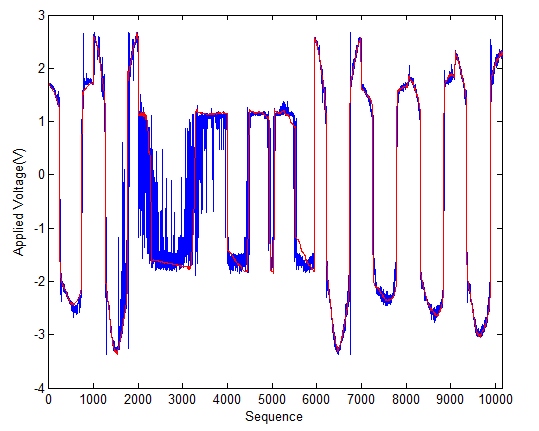}
\caption{Neural network output (blue) against training data set (red).}
\end{subfigure}
\begin{subfigure}{0.45\textwidth}
\includegraphics[width=\textwidth]{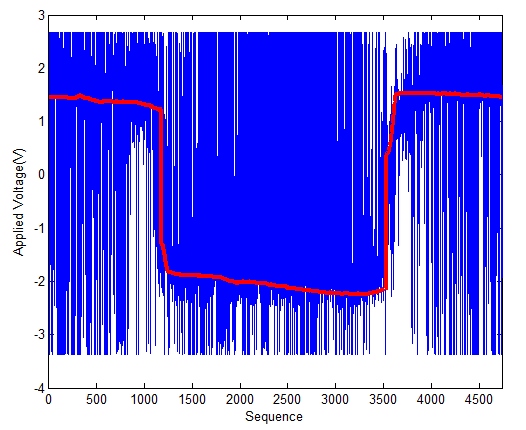}
\caption{Neural network output (blue) against test data set (red).}
\end{subfigure}
\caption{Performance of the trained network constructed using the typical approach (velocity and acceleration as input and applied voltage as output).}
\label{fig:typicalNNperformance}
\end{figure}

Figures \ref{fig:typicalNNperformance}a and \ref{fig:typicalNNperformance}b illustrate the performance of the neural network (after training for $>110$ epochs) against the training and test datasets respectively. The graphs plot control voltage (vertical axis) against the corresponding velocity and acceleration values (which are labelled by sequence index on the horizontal axis). The desired control voltage values generated from the training and test datasets are shown on their respective graphs in red. The actual control voltage values output by the neural network are shown in blue. It is apparent that while the neural network's output matches the training data fairly well, its performance against the test data is very poor.

\section{Neural Network Modelling -- Dominant Variable Approach}
\label{sect:NNDominantVariable}
It was found that the poor predictive performance of the preceding neural network inverse model could be rectified by restricting its input to only the dominant input variable of the inverse process. By ‘dominant input variable’ it is implied that the variable to which the output is most sensitive. To determine which input variable (i.e. velocity or acceleration) is dominant, it is necessary to develop an approximate analytical model of the piezoelectric positioning stage dynamics.

\subsection{Approximate Analytical Model of the Piezoelectric Positioning Stage}
Despite the complexity of the mechanism, with a little physical intuition it is possible to obtain a rough analytical model that suffices the purpose of identifying the dominant variable. Considering that the piezoelectric positioning stage consists of a platform that slides on rigid rails, friction is expected to play a major role in the disturbances affecting the system’s performance. Classical models of friction are constructed by various combinations of Coulomb, viscous, and drag friction as their basic components. A brief review of the various types of frictional forces and their mathematical representation is as follows:

\subsubsection{Frictional Effects}
\paragraph{Coulomb Friction}
Coulomb friction is a type of mechanical damping in which energy is consumed during sliding motion. The frictional force resulting from the relative motion of two surfaces in mutual contact always acts opposite to the relative motion and is proportional to the normal force of contact. It is simply given as
\begin{equation}
	F_c = f_c \sgn(\dot{x})
\end{equation}
where $f_c$ is the normal force.

\paragraph{Viscous Friction}
Viscous friction arises from the motion of any object through or past another object, and acts against this motion. Under well-lubricated conditions, the viscous frictional force is approximately proportional to velocity. It is characterised by the following linear relationship given as
\begin{equation}
	F_v = f_v \dot{x}
\end{equation}
where $f_v$ is the coefficient of viscous friction.

\paragraph{Drag Friction}
Drag friction is the frictional force on a solid object moving through a fluid. It is proportional to the square of the velocity given by
\begin{equation}
	F_d = f_d \dot{x} \abs{\dot{x}}
\end{equation}
where $f_d$ is the drag coefficient.

As the results of Section~\ref{sect:nonlinearCharacteristics} suggest, viscous and Coulomb friction effects are significant and should be accounted for in the model. On the other hand, drag friction can be neglected due to the low speed of motion of the stage and its low drag coefficient in air. 

\subsubsection{Parameter Identification}
Besides friction, the other major force acting on the piezoelectric position stage is the force provided by the actuator. On the basis of the manufacturer’s specifications, this is assumed to be a linear function of the applied voltage. Therefore, according to Newton’s second law the dynamics of the piezoelectric positioning stage can be described by the following second-order differential equation given as
\begin{equation} 
	\label{eqn:analyticalModel}
	\ddot{x} = -\frac{k_1}{m}\dot{x} - \frac{k_2}{m}\sgn(\dot{x}) + \frac{k_3}{m}u
\end{equation}
where $m$ denotes the rotor mass of piezoelectric positioning stage, $k_1$ and $k_2$ are respectively the viscous and Coulomb frictional coefficients, $u$ is the input voltage and $k_3$ is the constant relating voltage to actuator force. 

For notational convenience, and to emphasize that the frictional coefficients depend on the direction of motion (c.f. Section~\ref{sect:nonlinearCharacteristics}), re-write the coefficients in this manner as
\begin{align}
	a_1 = \frac{k_1}{m} = 
									\begin{cases}
										a_{1p}&	\dot{x} > 0 \\
										a_{1n}&	\dot{x} < 0	
									\end{cases} \qquad
	a_2 = \frac{k_2}{m} =
									\begin{cases}
										a_{2p}&	\dot{x} > 0 \\
										a_{2n}&	\dot{x} < 0	
									\end{cases} \qquad
	a_3 = \frac{k_3}{m}
\end{align}
According to manufacturer specifications, $a_3 = 6 \frac{N}{V \cdot kg}$. 
To obtain the values of the remaining parameters in \eqref{eqn:analyticalModel}, the stage is driven in both directions by subjecting it to a variety of positive and negative voltage pulses of duration 0.4s.

\begin{figure}
\centering
\begin{subfigure}{0.45\textwidth}
\includegraphics[width=\textwidth]{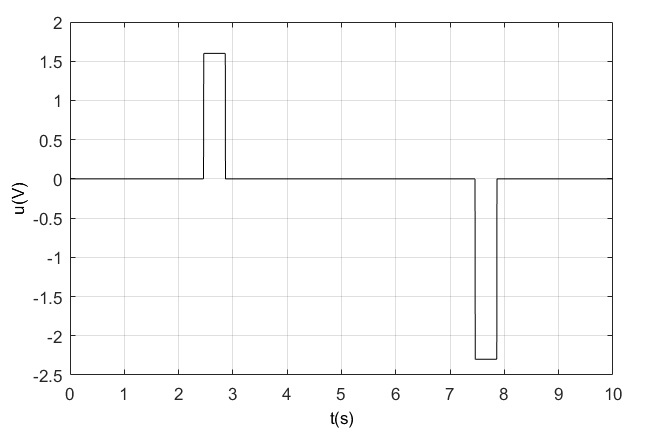}
\caption{Pulses applied.}
\label{fig:shortPulseInput}
\end{subfigure}
\begin{subfigure}{0.48\textwidth}
\includegraphics[width=\textwidth]{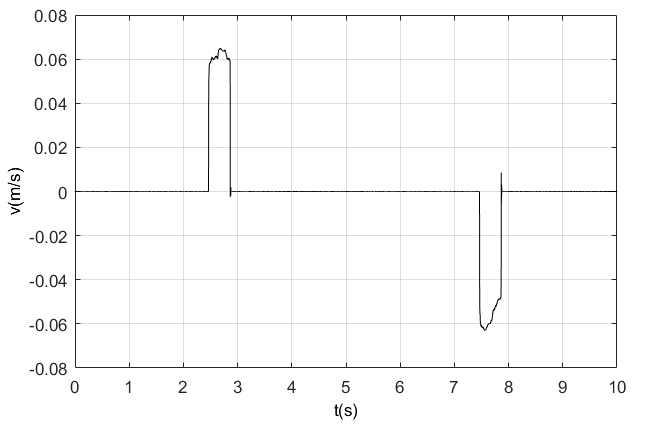}
\caption{Velocity of the positioning stage.}
\label{fig:shortPulseResponse}
\end{subfigure}
\caption{Application of voltage pulses 0.4s duration, with amplitudes of 1.6V and -2.3V respectively, and resulting velocity of the positioning stage.}
\label{fig:shortPulseInputResponse}
\end{figure}

Applying pulses of 1.6V and -2.3V amplitude and recording the resulting velocity of the positioning stage as shown in Fig.~\ref{fig:shortPulseInputResponse}, the following expressions are obtained that
\begin{equation}
\begin{cases}
0.05562 a_{1n} + a_{2n} - 2.3 \times 6 = 0	\\
-0.06222 a_{1p} - a_{2p} + 1.6 \times 6 = 0
\end{cases}
\end{equation}

Following a similar method, using pulses of similar duration ($0.4s$) and various amplitudes of $-1.8V$, $1.3V$, $-2V$, $1.5V$, $-2.1V$, $1.7V$, $-2.5V$, $2V$ respectively, it is obtained that

\begin{align}
&\begin{cases}
0.03393 a_{1n} + a_{2n} - 1.8 \times 6 = 0	\\
-0.04465 a_{1p} - a_{2p} + 1.3 \times 6 = 0
\end{cases}\\
&\begin{cases}
0.04622 a_{1n} + a_{2n} - 2.0 \times 6 = 0	\\
-0.05742 a_{1p} - a_{2p} + 1.5 \times 6 = 0
\end{cases}\\
&\begin{cases}
0.04991 a_{1n} + a_{2n} - 2.1 \times 6 = 0	\\
-0.06863 a_{1p} - a_{2p} + 1.7 \times 6 = 0
\end{cases}\\
&\begin{cases}
0.07120 a_{1n} + a_{2n} - 2.5 \times 6 = 0	\\
-0.08519a_{1p} - a_{2p} + 2.0 \times 6 = 0
\end{cases}
\end{align}

Using the method of Least-Squares, let 
\begin{align}
\vec{A} &= \begin{bmatrix} a_{1p} & a_{1n} & a_{2p} & a_{2n} \end{bmatrix}^T \\
\vec{Y} &= \begin{bmatrix}10.8 & 12 & 12.6 & 13.8 & 15 & -7.8 & -9 & -9.6 & -10.2 & -12\end{bmatrix}^T 
\end{align}
and $\vec{X}$ be the coefficients of $\vec{A}$, and solve the equation \eqref{eqn:leastSquares} as

\begin{equation}
	\label{eqn:leastSquares}
	\vec{A} = (\vec{X}^T \vec{X})^{-1} \vec{X}^T \vec{Y}
\end{equation}

Thus the values for $a_{1p}, a_{1n}, a_{2p}, a_{2n}$ are obtained as
\begin{align}
&\begin{cases}
a_{1p} &= 104.0154\\
a_{1n} &= 117.1441
\end{cases}
&\begin{cases}
a_{2p} &= 3.1023\\
a_{2n} &= 6.8216
\end{cases}
\end{align}

\begin{figure}
\centering
\begin{subfigure}{0.45\textwidth}
\includegraphics[width=\textwidth]{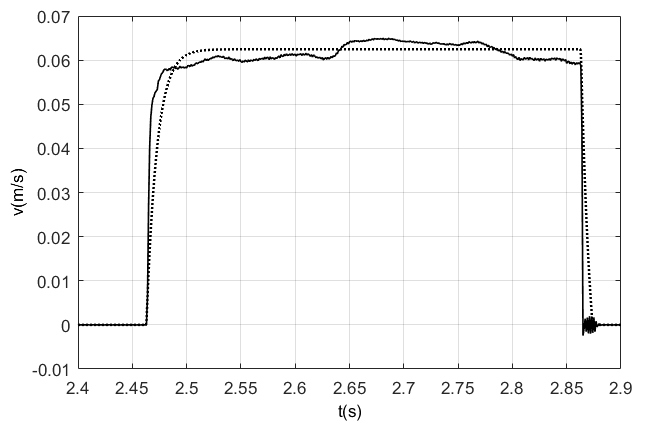}
\end{subfigure}
\begin{subfigure}{0.45\textwidth}
\includegraphics[width=\textwidth]{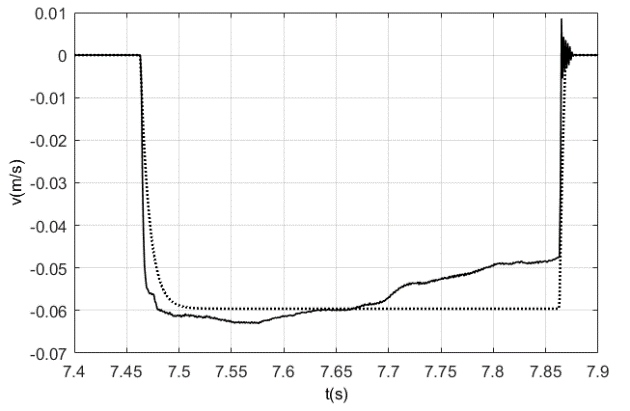}
\end{subfigure}
\caption{Comparison of the velocity computed using (dotted line) with the actual response of the system (solid line), in response to input pulses of 0.4s duration, with amplitudes of 1.6V and -2.3V respectively}
\label{fig:velocityVsResponse}
\end{figure}

To verify these values, they were substituted into \eqref{eqn:analyticalModel} to compute the velocity of the positioning stage in response to applied voltage pulses of amplitude $1.6V$ and $-2.3V$. Fig.~\ref{fig:velocityVsResponse} shows a comparison of the velocity computed from the model (dotted line) to the actual response of the system (solid line).

\begin{figure}
\centering
\begin{subfigure}{0.45\textwidth}
\includegraphics[width=\textwidth]{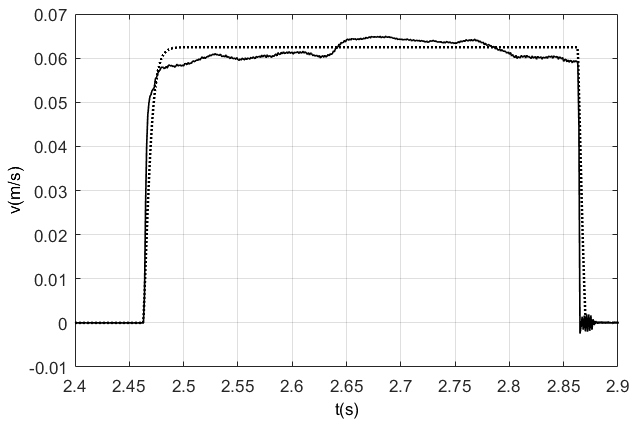}
\end{subfigure}
\begin{subfigure}{0.45\textwidth}
\includegraphics[width=\textwidth]{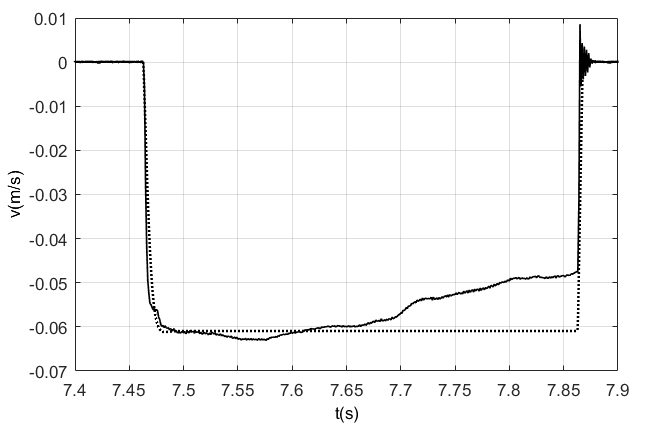}
\end{subfigure}
\caption{Comparison of the velocity computed using the model (dotted line) with the actual response of the system (solid line), under the same conditions as those for which Fig.~\ref{fig:velocityVsResponse} are obtained, after adjustments to the viscous friction term.}
\label{fig:velocityVsResponse2}
\end{figure}

From Fig.~\ref{fig:velocityVsResponse}, it is seen that the frictional force in \eqref{eqn:analyticalModel} appears delayed by about $0.0035s$ compared to the actual behaviour. The viscous friction term in \eqref{eqn:analyticalModel} is adjusted to correct for this, in order to achieve a better match between the modelled and actual velocities, as shown in Fig.~\ref{fig:velocityVsResponse2}. Therefore, the dynamical model is obtained as

\begin{equation} 
\label{eqn:analyticalModelFinal}
	\ddot{x}(t) = -a_1 \dot{x}(t - 0.0035) - a_2 \sgn(\dot{x}(t)) + 6u(t)
\end{equation}
where
\begin{equation}
	a_1 = \begin{cases}
				104.0154&	\dot{x}(t-0.0035) > 0 \\
				117.1441&	\dot{x}(t-0.0035) < 0
			\end{cases} \qquad
	a_2 = \begin{cases}
				3.1023&	\dot{x}(t) > 0 \\
				6.8216&	\dot{x}(t) < 0 
			\end{cases}
\end{equation}

To verify this model, a comparison is made between the simulated and actual velocities of the piezoelectric stage in response to a triangular function input with a period of 6 seconds and peak amplitudes at 1.5V and -2.1V (see Fig. 7, dotted line shows velocity computed with the model, solid line shows velocity measured from actual system). It can be seen that although the model is not accurate, the results are in qualitative agreement with the actual system, and this is sufficient for the purpose of identifying the dominant variable.

\begin{figure}
\centering
\includegraphics[width=0.7\textwidth]{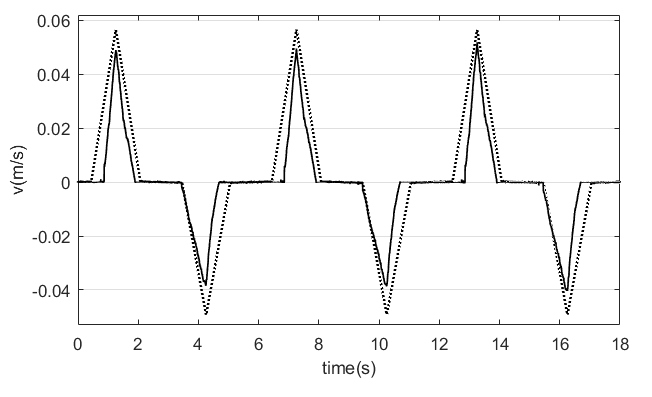}
\caption{Comparison of the velocity computed using the model (dotted line) with the actual response of the system (solid line), in response to a triangle function input of 6 second period and peak amplitudes at 1.5V and -2.1V.}
\label{fig:analyticalModelVsMeasuredTriangleInput}
\end{figure}

\subsection{Identification of the Dominant Variable}
Rearranging \eqref{eqn:analyticalModelFinal} to write the applied voltage $u$ in terms of the velocity and acceleration it is obtained that
\begin{equation}
	6u(t) = \ddot{x}(t) + a_1 \dot{x}(t - 0.0035) + a_2 \sgn(\dot{x}(t))
\end{equation}
The variable giving rise to terms which together make the largest contribution to the magnitude of $u$ is considered the dominant variable. For this purpose it suffices to compare the term due to acceleration $\ddot{x}(t)$ with the term due to velocity $a_1 \dot{x}(t)$ (the Coulomb friction term can be neglected because it is effectively a constant).

\begin{figure}
\centering
\includegraphics[width=0.7\textwidth]{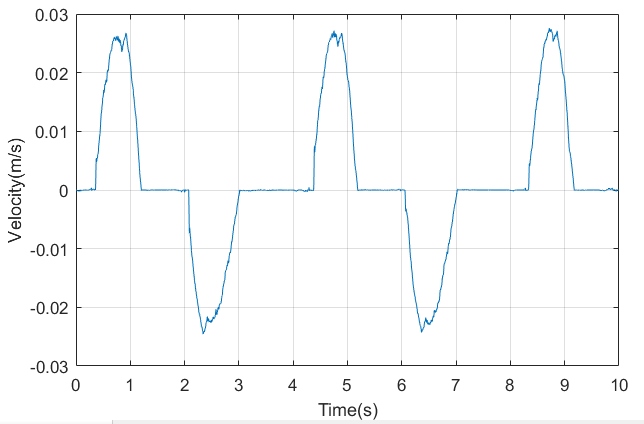}
\caption{Velocity of the piezoelectric positioning stage in response to a sinusoidal input voltage with amplitude 1.5V and frequency 0.25 Hz.} 
\label{fig:velocityRespToSinusoidInput}
\end{figure}

\begin{figure}
\centering
\includegraphics[width=0.7\textwidth]{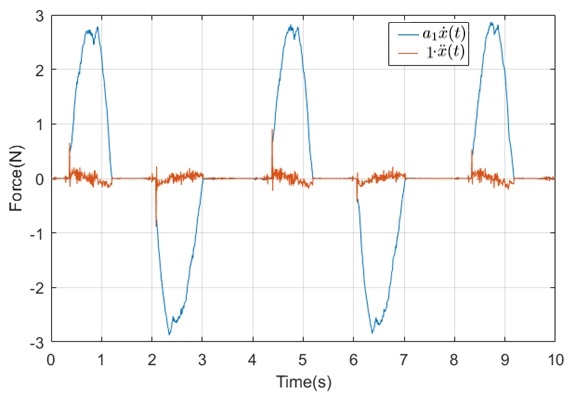}
\caption{Comparison of the terms $\ddot{x}(t)$ and $a_1 \dot{x}(t)$.}
\label{fig:dominantVariableComparison}
\end{figure}

These terms can be evaluated by applying an input signal to the piezoelectric positioning stage and recording the resulting velocity. Fig~\ref{fig:velocityRespToSinusoidInput} shows the velocity resulting from application of the signal 
$u = 1.5 \sin (\frac{\pi}{2}t - \frac{\pi}{2}) - 0.3$.
From this, the value of the terms $\ddot{x}(t)$ and $a_1 \dot{x}(t)$ are computed and compared (Fig~\ref{fig:dominantVariableComparison}).

From Fig~\ref{fig:dominantVariableComparison} the bound of $\ddot{x}(t) / a_1 \dot{x}(t)$ is found to be $O(10^{-1})$, hence the output of the inverse system is dominated by its velocity. 

\subsection{Setting the Size of the Hidden Layer}
The neural network constructed with the new approach comprises a single neuron each in the input and output layers. The input is the velocity of the positioning stage (which was identified as the dominant variable), and the output is the applied voltage. It is however, not straightforward to determine the appropriate number of neurons for the hidden layer, with respect to the performance of the network.
The mean squared error (MSE) between the actual output $y_d$ and predicted output $y_p$ is used as the criterion for checking convergence of networks, and can be calculated with the equation given as
\begin{equation}
	E_{MSE} = \frac{1}{q} \sum_{i=1}^{q} (y_{di} - y_{pi})^2
\end{equation}
where $q$ is the number of patterns presented in a data set.
Using this measure, the effect of the number of hidden layer neurons on the network performance was experimentally investigated, and results are shown in Table~\ref{tab:hiddenNeurons}.

\begin{table}[h] 
\caption{Effect of hidden-layer neurons on the network performance}
\centering
\begin{tabular}{p{2.5cm}p{2cm}p{2cm}p{2cm}}  
\toprule
Hidden Layer Neurons    & Training Epochs & Training $E_{MSE}$ & Testing $E_{MSE}$ \\
\midrule
16	&	335	&	0.120776	&	0.054228 \\
18	&	327	&	0.120892	&	0.053226 \\
20	&	754	&	0.119935	&	0.053679 \\
22	&	679	&	0.118807	&	0.054072 \\
24	&	688	&	0.118161	&	0.052982 \\
26	&	374	&	0.119766	&	0.055169 \\
28	&	350	&	0.119977	&	0.053508 \\
30	&	369	&	0.120442	&	0.055209 \\
32	&	754	&	0.119834	&	0.054972 \\
\bottomrule
\end{tabular}
\label{tab:hiddenNeurons}
\end{table}

Table~\ref{tab:hiddenNeurons} indicates that increasing the number of neurons in the hidden layer does not simply reduce the average MSE. Rather, it appears that the optimal number of the neurons in the hidden layer is 24. The inverse model for the piezoelectric positioning stage is thus given by the neural network given as
\begin{equation}\label{eqn:finalNNmodel}
	y_p = \vec{W}_2^T f(\vec{W}_1^T v + \vec{B}_1) + b_2
\end{equation}
where $f$ is the hyperbolic tangent function, $v$ is the velocity, $y_p$ is applied voltage, and the connection weights and biases are given as
\begin{align}
\vec{W}_1 =& \nonumber\\
[&-35.3468 \quad	34.6366 \quad -28.435 \quad -33.44 \quad -33.4596 \quad 32.1285 \nonumber\\
&-29.9843 \quad 43.3091 \quad 	4.3072 \quad -136.8754 \quad -231.9848   \quad 239.7699 \nonumber\\  
&-274.3949  \quad 225.5036 \quad  -224.9426 \quad  -5.3267  \quad 44.2194 \quad  65.8648 \nonumber\\  
&33.4028 \quad  -48.4009 \quad -22.5532 \quad  -24.6959  \quad -38.4669 \quad  36.3986]^T \nonumber\\
\end{align}
\begin{align}
\vec{B}_1 =& \nonumber\\
[&33.727 \quad -31.4364 \quad 23.0972 \quad 24.6802 \quad 22.7881 \quad -19.5757 \nonumber\\ &15.8065 \quad -17.4002 \quad -0.49335 \quad 7.0838 \quad 0.00042311 \quad 0.27504 \quad \nonumber\\
&-0.40677 \quad 0.84943 \quad -10.6875 \quad -1.408 \quad 17.2613 \quad 33.2112 \nonumber\\ &29.6669 \quad -28.3202 \quad -15.6053 \quad -19.8161 \quad -34.0213 \quad 33.5377]^T \nonumber\\
\end{align}
\begin{align}
\vec{W}_2 =& \nonumber\\
&[-0.16523 \quad 0.10884 \quad -0.1075 \quad -0.034101 \quad -0.051026 \quad 0.024077 \nonumber\\ 
&-0.074166 \quad 0.037539 \quad 0.1368 \quad -0.066009 \quad -32.9146 \quad -85.532 \nonumber\\ &-44.7181 \quad 8.3888 \quad -0.10717 \quad -0.06488 \quad 0.040388 \quad 0.041598 \nonumber\\ &-3.6387 \quad -0.035379 \quad -0.10142 \quad -0.13976 \quad -3.1364 \quad 0.75016]^T\nonumber\\
\\
\nonumber\\
b_2 &=~ 0.0072113
\end{align}

\begin{figure}
\centering
\includegraphics[width=0.7\textwidth]{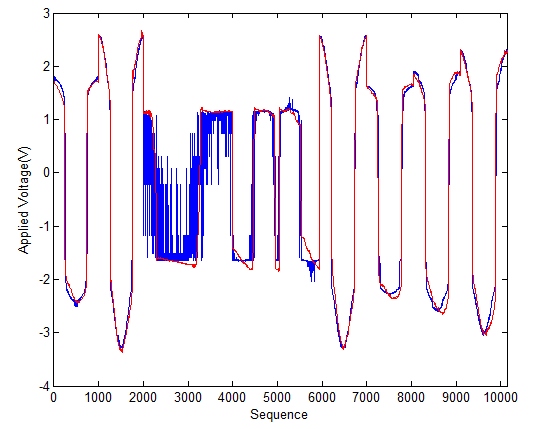}
\caption{Performance of the trained network constructed using the dominant variable approach (Eqn. \eqref{eqn:finalNNmodel}). (Neural network output in blue, training data set in red.)}
\label{fig:dominantVarNNperformance}
\end{figure}

The performance of the neural network against the training data set after 688 epochs (c.f. Table~\ref{tab:hiddenNeurons}) is shown in Fig.~\ref{fig:dominantVarNNperformance}.

\section{Prediction Performance of the Neural Network Model Based on the Dominant Variable Approach}
\label{sect:predictionPerformance}
The prediction performance of the neural network was evaluated in two ways, first by checking its output against measured behaviour (i.e. with the test data set), and second by its effectiveness in providing feedforward compensation to a PI controller for position tracking. The results are as follows:

\subsection{Prediction Performance with Test Data Set}
\begin{figure}
\centering
\begin{subfigure}{0.45\textwidth}
\includegraphics[width=\textwidth]{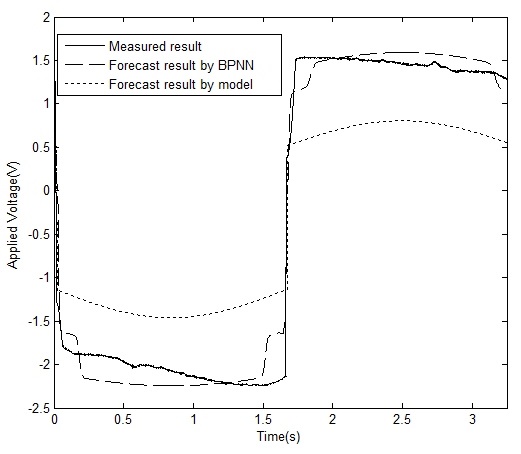}
\caption{$v = 0.016625 \sin(1.9t)$}
\end{subfigure}
\begin{subfigure}{0.45\textwidth}
\includegraphics[width=\textwidth]{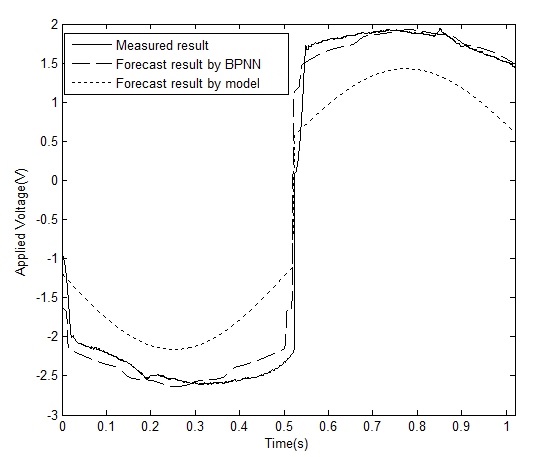}
\caption{$v=0.0525\sin(6t)$}
\end{subfigure}
\begin{subfigure}{0.45\textwidth}
\includegraphics[width=\textwidth]{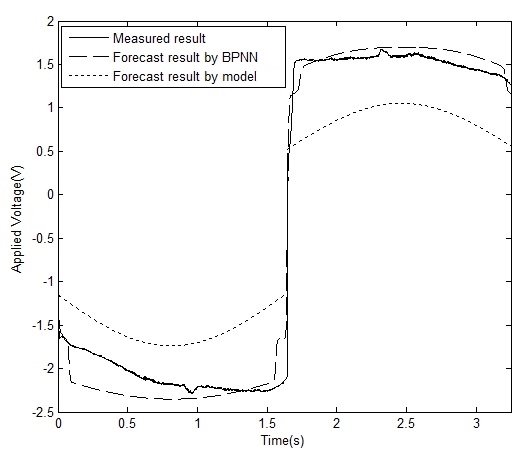}
\caption{$v=0.030875\sin(1.9t)$}
\end{subfigure}
\begin{subfigure}{0.45\textwidth}
\includegraphics[width=\textwidth]{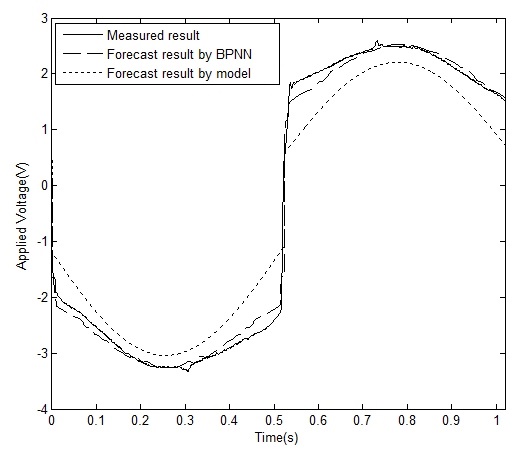}
\caption{$v=0.0975\sin(6t)$}
\end{subfigure}
\caption{Comparison of predictions of the neural network model using the dominant variable approach, and the analytical model, with the test data set.}
\label{fig:performanceComparisonTestData}
\end{figure}

Fig.~\ref{fig:performanceComparisonTestData} shows the outputs of the neural network \eqref{eqn:finalNNmodel} (dashed lines) and the approximate analytical model \eqref{eqn:analyticalModelFinal} (dotted lines) in response to each of the four signals comprising the test data set (c.f. Fig.~\ref{fig:NNdataSets}). The results indicate that the neural network's predictions are in fairly good agreement with actual behaviour, and are much more accurate than the analytical model.

\subsection{Prediction Performance with Other Input Waveforms}
\begin{figure}
\centering
\begin{subfigure}{0.45\textwidth}
\includegraphics[width=\textwidth]{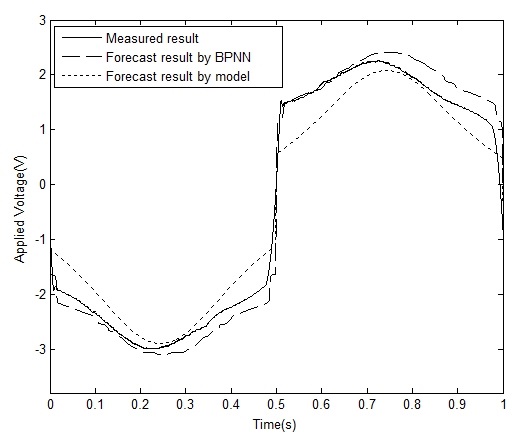}
\caption{$v = \frac{0.8}{\pi^2}[\cos(2 \pi t) + \frac{1}{9} \cos(6 \pi t)]$}
\end{subfigure}
\begin{subfigure}{0.45\textwidth}
\includegraphics[width=\textwidth]{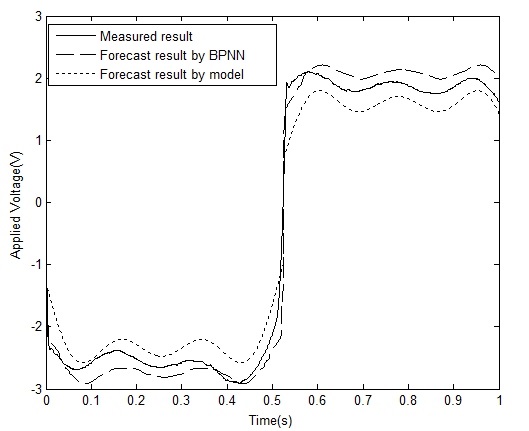}
\caption{$v = \frac{0.248}{\pi}[\cos(6t) - \frac{1}{3} \cos (18t) + \frac{1}{5} \cos(30t)]$}
\end{subfigure}
\caption{Further comparison of predictions of the neural network model using the dominant variable approach, and the analytical model.}
\label{fig:performanceComparisonOtherWaveforms}
\end{figure}

Prediction performance of the neural network was further validated by another set of data, generated in the same manner as described in section \ref{sect:trngPerformance}, but using approximate triangular and square position reference signals instead. (These signals were approximated with sinusoids in order to avoid exceeding the maximum permissible velocity of the piezoelectric stage when collecting measurement data.) Likewise, results in Fig.~\ref{fig:performanceComparisonOtherWaveforms} indicate better performance compared with the analytical model.

\subsection{Prediction Performance in Feedforward Compensation}
\begin{figure}
\centering
\includegraphics[width=0.9\textwidth]{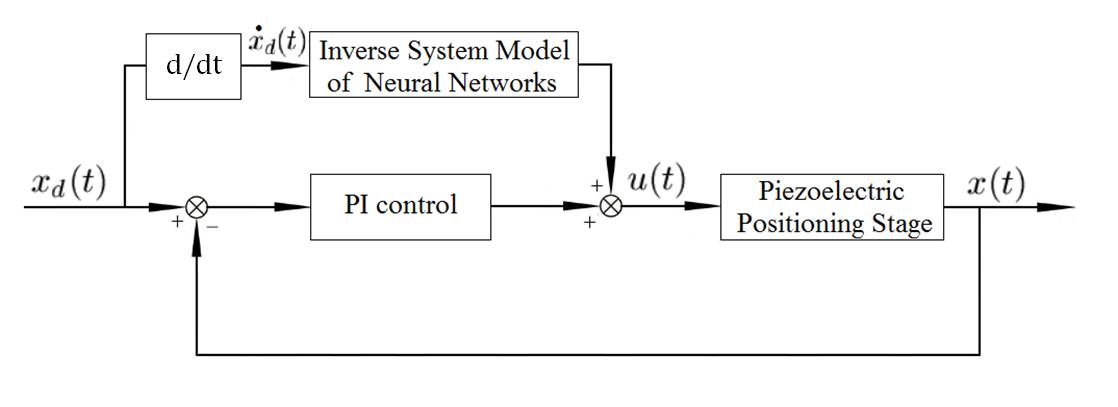}
\caption{Block diagram of PI position controller with neural network feedforward compensation.}
\label{fig:blockDiagram}
\end{figure}

\begin{figure}
\centering
\includegraphics[width=0.7\textwidth]{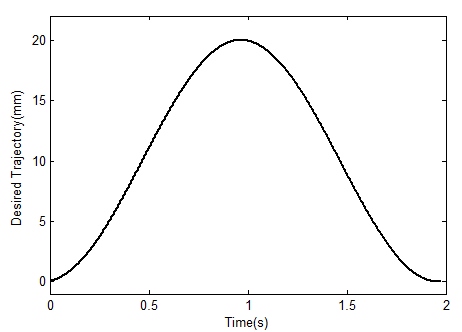}
\caption{Position reference signal to be tracked.}
\label{fig:posControlReference}
\end{figure}

\begin{figure}
\centering
\includegraphics[width=0.7\textwidth]{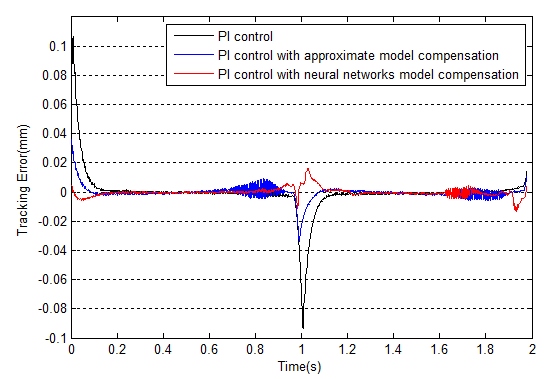}
\includegraphics[width=0.75\textwidth]{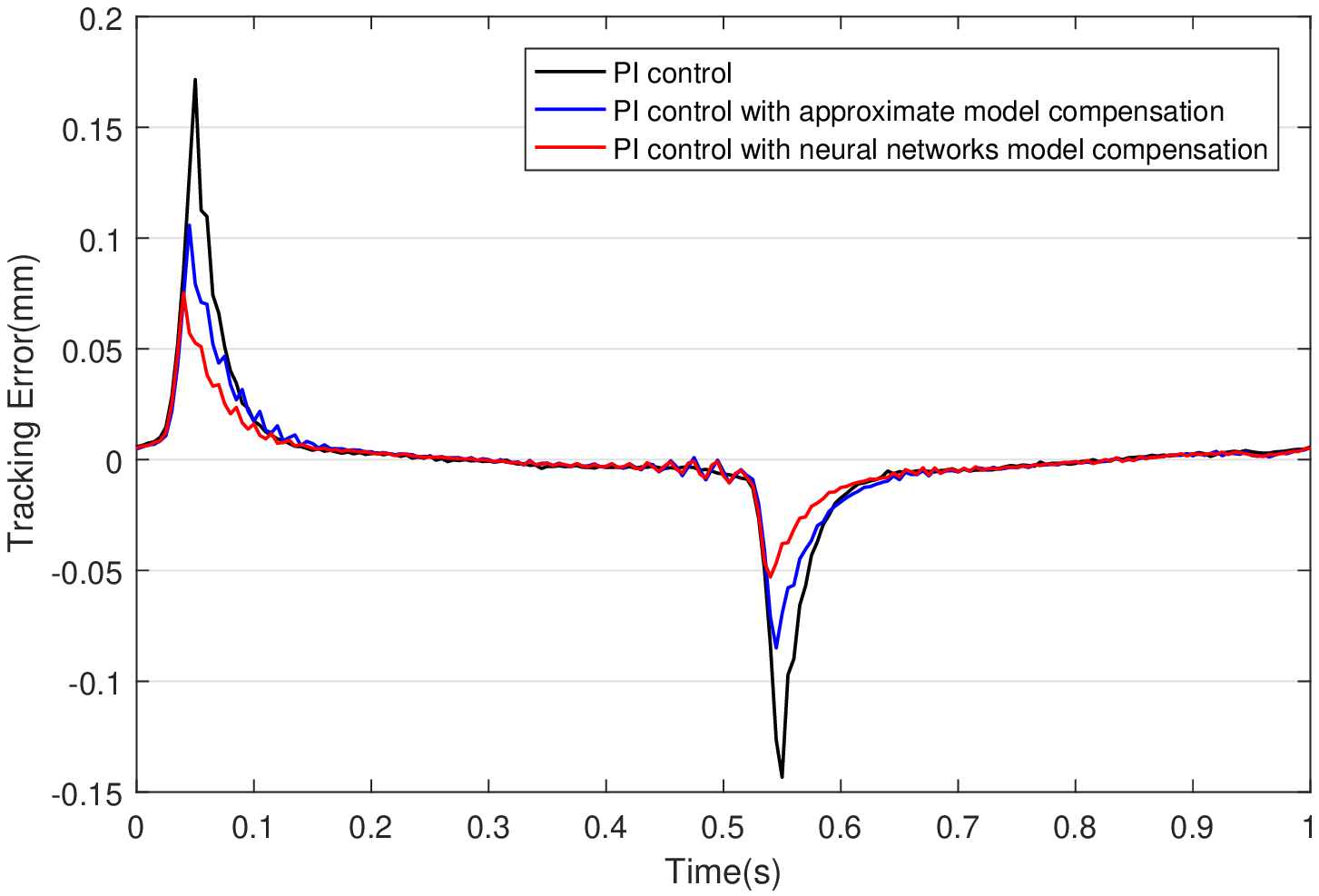}
\begin{tabular}{p{6cm}p{2cm}p{2cm}}  
\toprule
Control Method & \multicolumn{2}{ c }{Maximum Error (mm)} \\
				&	at 0.5Hz		& at 1.0Hz \\
\midrule
PI controller	&	 	0.10651		&	0.17156 \\
PI controller with approximate model compensation	 &	0.03609		&	0.10579\\
PI controller with neural network model compensation 	&	0.01779	 &	0.07292\\
\bottomrule
\end{tabular}
\caption{Comparisons of the tracking error of the controller using various compensation schemes, at different reference signal frequencies (0.5Hz at top, 1.0Hz at bottom).}
\label{fig:posControlError}
\end{figure}

\begin{figure}
\centering
\includegraphics[width=0.7\textwidth]{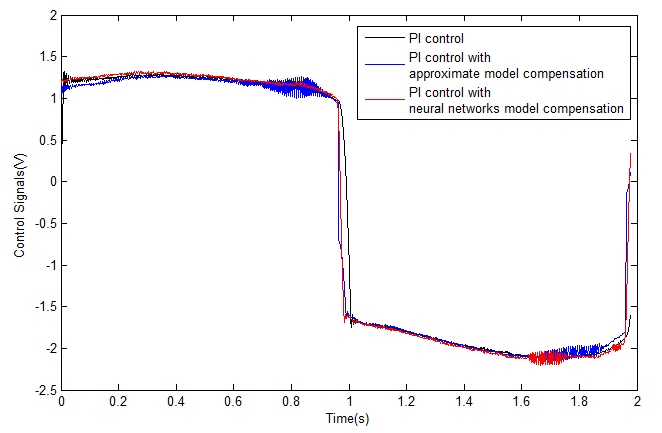}
\includegraphics[width=0.8\textwidth]{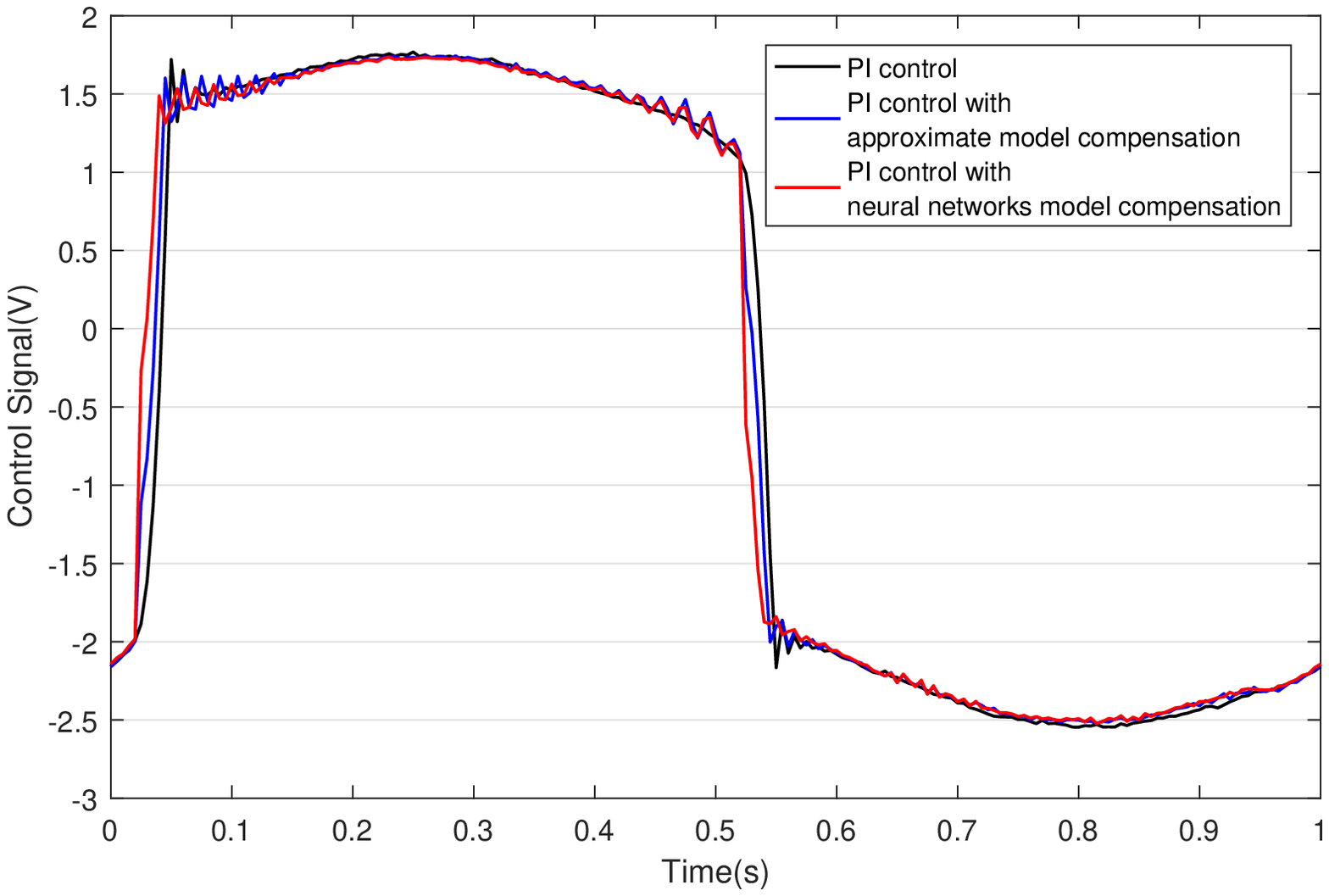}
\caption{Comparisons of the control signals generated by the controller using various compensation schemes, at different reference signal frequencies (0.5Hz at top, 1.0Hz at bottom).}
\label{fig:posControlSignals}
\end{figure}

The effectiveness of the inverse neural network model for feedforward compensation was evaluated by comparing the position tracking performance of a PI controller alone, and with feedforward compensation using either the approximate analytical model or the neural network (see Fig.~\ref{fig:blockDiagram} for block diagram). \textit{The position signal to be tracked is shown in Fig.~\ref{fig:posControlReference} and a smilar signal at a higher frequency of 1.0Hz is also used.} PI controller gains were adjusted (in the absence of feedforward compensation) in order to achieve minimal tracking error without oscillation, resulting in a proportional gain of 19000 and an integral gain of 660000. A comparison of the tracking errors and control signals generated by the PI controller alone, and with each type of feedforward compensation, is shown in Figures \ref{fig:posControlError} and \ref{fig:posControlSignals} respectively.

As was discussed earlier, for the piezoelectric position stage, the frictional disturbance influence the dyanamics of the system greatly and, therefore, it is expected that the tracking would suffer at low speeds. From Fig.\ref{fig:posControlError} it is seen that error is greatest at the onset of motion and attenuates once the stage is in motion. Even though integral action is capable in attenuating the effects of the disturbance the PI controller on its own is unable to react quickly enough. As it would be expected, the performance with feedforward compensation would be better because of the instantaneous compensation of the frictional disturbance. The results of Fig.~\ref{fig:posControlError} demonstrate that use of the neural network inverse model for feedforward compensation can greatly improve the position control accuracy of the piezoelectric positioning stage because of the higher accuracy in the model compared to the approximate model.

\section{Conclusions}
An approach for developing a neural network inverse model of a piezoelectric positioning stage, which exhibits rate-dependent, asymmetric hysteresis has been presented. The difficulties in modeling piezoelectric actuators has been discussed and various approaches has been compared to the proposed approach. It was shown that the dominant variable approach to neural network design is capable of producing better prediction results as compared to classical approaches. Furthermore, use of this dominant variable inverse model as a feedforward compensator in conjunction with a PI position controller demonstrated much better tracking performance with lower error compared to compensation using the approximate analytical model. 





\end{document}